\documentstyle[pra,aps]{revtex}
\begin{document}
\draft
\title{Teleportation of rotations and receiver-encoded secret sharing}
\author{Chui-Ping Yang and Julio Gea-Banacloche\thanks{%
Email address: jgeabana@uark.edu}}
\address{Department of Physics, University of Arkansas, Fayetteville, Arkansas 72701}
\maketitle

\begin{abstract}
We show how an arbitrary qubit rotation can be teleported, albeit probabilistically,
using 1 e-bit of entanglement and one classical bit.
We use this to present a scheme for
implementing quantum secret sharing. The scheme operates essentially by
sending a ``secret" rotated qubit of information to several users, who need
to cooperate in order to recover the original qubit.
\end{abstract}


\date{\today}


\begin{center}
{\bf I. INTRODUCTION}
\end{center}

Hiding information may be one of the most useful applications of the growing
science of quantum information, beginning with the classical quantum
cryptography work of Bennett et al. [1-2]. Since that work, many other
cryptographic problems have been addressed in a quantum context. We may
cite, as especially relevant to this paper, the work of Hillery et al. [3]
(see also Ref. [4]) and Cleve et al. on quantum secret sharing [5], and the
very recent work by Terhal et al. [6]. One of the problems considered in [3] and [4]
was how one party (Alice) could send a
qubit of (quantum) information to two agents, Bob and Charlie, in such a way
that they would have to cooperate in order to recover the original message.
Cleve et al. addressed the general problem of hiding the state of a ($d$%
-dimensional, with $d$ arbitrary) quantum system, by encoding it into $n$
shares, in such a way that $k$ shares would be necessary to recover the
secret, and $k-1$ shares would contain no information whatever (a $\left(
k,n\right) $ $threshold$ $\ scheme$).

In this paper, we present a scheme which allows $n$ potential receivers of a 
qubit of quantum information to manipulate, from a distance, its state (albeit 
without learning
anything about it), in such a way that the original information becomes
``hidden'' (either completely or only partly, depending on the qubit's initial 
state); then the qubit may be sent to one of them, and all of them must 
collaborate, by exchanging classical information, in order to recover the
full original state.  If even one of them does not collaborate, the best they
can do is to leave the qubit in a random rotated state.

Our method makes use of a maximally entangled 
(GHZ) state which Alice (the sender) shares with all the receivers.  We show
here that such a state allows the receivers to remotely ``rotate'' Alice's 
qubit; specifically, we show that an arbitrary rotation about the ``y'' axis
can be performed remotely, albeit probabilistically, at the cost of only one e-bit of 
entanglement and one bit of classical communication.  It should be noted that Huelga et al. [7] 
first showed that the
teleportation of an arbitrary unitary operation requires a minimum of 2
e-bits and 2 classical bits, and more recently [8] they have also established the existence of
restricted sets of operations which require fewer resources in order to be
teleported, either probabilistically or deterministically.  Our result can be regarded
as an additional example of this kind.

The paper is organized as follows.  The method for the teleportation of multiple 
rotations using a GHZ state is presented in the following Section (Section II).  The
possible applications to secret sharing, including some considerations about 
the security of the scheme, are presented in Section III.  Section IV contains a brief 
discussion and conclusions.

\begin{center}
{\bf II. TELEPORTATION OF ROTATIONS}
\end{center}

Suppose Alice holds a two-state particle (i.e., qubit), which is labeled by $%
a$ and in an arbitrary unknown pure state $\alpha \left| 0\right\rangle
_{a}+\beta \left| 1\right\rangle _{a}.$ We will show how $n$ distant users
can apply an arbitrary rotation to Alice's ``message'' qubit $a$ through
their local operations and classical communications. The ``rotation'' we have in mind is 
a formal rotation, by an angle $\theta$, in the two-dimensional Hilbert space 
of the qubit (see Eq.~(4) below); it is,
however, easy to show that the same operation corresponds, in Bloch-sphere terms,
to a rotation by an angle $2\theta$ around the $y$ axis, that is, to the action 
of the operator $\exp(-i\sigma_y\theta)$.

To begin with, Alice
needs to share a $\left( n+1\right) $ -qubit GHZ state with $n$ users [9]; the
GHZ qubit belonging to Alice is labeled by $b$, and the shared GHZ state is 
$\left( \left| 0\right\rangle _{b}\left| 00...0\right\rangle +\left|
1\right\rangle _{b}\left| 11...1\right\rangle \right) $. Thus, the state of
the whole system is 
\begin{equation}
\left( \alpha \left| 0\right\rangle _{a}+\beta \left| 1\right\rangle
_{a}\right) \otimes \left( \left| 0\right\rangle _{b}\left|
00...0\right\rangle +\left| 1\right\rangle _{b}\left| 11...1\right\rangle
\right) .
\end{equation}
By Alice first performing a Control-Rotation operation $R_{ab}$ [10] on her
qubits $a$ and $b$ (with the control being qubit $a$) 
\begin{equation}
R_{ab}=\left| 00\right\rangle \left\langle 00\right| +\left| 01\right\rangle
\left\langle 01\right| +\left| 10\right\rangle \left\langle 10\right|
-\left| 11\right\rangle \left\langle 11\right| ,
\end{equation}
and then a Control-Not operation $C_{ab}$, Eq. (1) will be transformed into 
\begin{equation}
\left[ \left( \alpha \left| 00\right\rangle _{ab}+\beta \left|
11\right\rangle _{ab}\right) \left| 00...0\right\rangle +\left( \alpha
\left| 01\right\rangle _{ab}-\beta \left| 10\right\rangle _{ab}\right)
\left| 11...1\right\rangle \right] .
\end{equation}
Now each user performs a rotation operation on his/her qubit 
\begin{eqnarray}
\left| 0\right\rangle _{i} &\rightarrow &R\left( \theta _{i}\right) \left|
0\right\rangle _{i}=\cos \theta _{i}\left| 0\right\rangle _{i}+\sin \theta
_{i}\left| 1\right\rangle _{i},  \nonumber \\
\left| 1\right\rangle _{i} &\rightarrow &R\left( \theta _{i}\right) \left|
1\right\rangle _{i}=-\sin \theta _{i}\left| 0\right\rangle _{i}+\cos \theta
_{i}\left| 1\right\rangle _{i}
\end{eqnarray}
where subscript $i$ stands for the $i$th user and $\theta _{i}$ stands for
the $i$th user's rotation angle. After that, we get, from Eq. (3), 
\begin{eqnarray}
&&\left( \alpha \left| 00\right\rangle _{ab}+\beta \left| 11\right\rangle
_{ab}\right) \prod_{i=1}^{n}\left( \cos \theta _{i}\left| 0\right\rangle
_{i}+\sin \theta _{i}\left| 1\right\rangle _{i}\right)   \nonumber \\
&&+\left( \alpha \left| 01\right\rangle _{ab}-\beta \left| 10\right\rangle
_{ab}\right) \prod_{i=1}^{n}\left( -\sin \theta _{i}\left| 0\right\rangle
_{i}+\cos \theta _{i}\left| 1\right\rangle _{i}\right) 
\end{eqnarray}
Then, each user performs a measurement on his/her qubit.  Suppose that $m$
users (for simplicity, let them be the users labeled by 1, 2, ..., $m)$
measure their qubits in the $\left| 0\right\rangle $ while $n-m$ users
measure their qubits in the $\left| 1\right\rangle $. From Eq. (5), we have 
\begin{eqnarray}
&&\left( \alpha \left| 00\right\rangle _{ab}+\beta \left| 11\right\rangle
_{ab}\right) \prod_{i=1}^{m}\cos \theta _{i}\prod_{i=m+1}^{n}\sin \theta _{i}
\nonumber \\
&&+\left( \alpha \left| 01\right\rangle _{ab}-\beta \left| 10\right\rangle
_{ab}\right) \left( -1\right) ^{m}\prod_{i=1}^{m}\sin \theta
_{i}\prod_{i=m+1}^{n}\cos \theta _{i}
\end{eqnarray}
A simple SWAP operation on the qubits $a$ and $b$ by Alice will transform
Eq. (6) as follows 
\begin{eqnarray}
&&\left( \alpha \left| 00\right\rangle _{ab}+\beta \left| 11\right\rangle
_{ab}\right) \prod_{i=1}^{m}\cos \theta _{i}\prod_{i=m+1}^{n}\sin \theta _{i}
\nonumber \\
&&+\left( \alpha \left| 10\right\rangle _{ab}-\beta \left| 01\right\rangle
_{ab}\right) \left( -1\right) ^{m}\prod_{i=1}^{m}\sin \theta
_{i}\prod_{i=m+1}^{n}\cos \theta _{i}
\end{eqnarray}
The above equation (7) can be rewritten as 
\begin{equation}
\alpha \left| 0\right\rangle _{b}\left| \psi \right\rangle +\beta \left|
1\right\rangle _{b}\left| \psi ^{\prime }\right\rangle 
\end{equation}
where 
\begin{eqnarray}
\left| \psi \right\rangle  &=&\prod_{i=1}^{m}\cos \theta
_{i}\prod_{i=m+1}^{n}\sin \theta _{i}\left| 0\right\rangle _{a}+\left(
-1\right) ^{m}\prod_{i=1}^{m}\sin \theta _{i}\prod_{i=m+1}^{n}\cos \theta
_{i}\left| 1\right\rangle _{a}, \\
\left| \psi ^{\prime }\right\rangle  &=&\prod_{i=1}^{m}\cos \theta
_{i}\prod_{i=m+1}^{n}\sin \theta _{i}\left| 1\right\rangle _{a}-\left(
-1\right) ^{m}\prod_{i=1}^{m}\sin \theta _{i}\prod_{i=m+1}^{n}\cos \theta
_{i}\left| 0\right\rangle _{a}
\end{eqnarray}
Now Alice performs a Hadamard transform on her qubit $b:$ $\left|
0\right\rangle \rightarrow \left( \left| 0\right\rangle +\left|
1\right\rangle \right) $ and $\left| 1\right\rangle \rightarrow \left(
\left| 0\right\rangle -\left| 1\right\rangle \right) .$ After that, Eq. (8)
will be 
\begin{equation}
\left[ \alpha \left( \left| 0\right\rangle _{b}+\left| 1\right\rangle
_{b}\right) \left| \psi \right\rangle +\beta \left( \left| 0\right\rangle
_{b}-\left| 1\right\rangle _{b}\right) \left| \psi ^{\prime }\right\rangle %
\right] 
\end{equation}
One can easily find from Eq. (11) that if Alice performs a measurement on
her qubit $b$, for the measurement outcomes $\left| 0\right\rangle $ and $%
\left| 1\right\rangle $, Eq. (11) will be, respectively 
\begin{eqnarray}
\left| 0\right\rangle _{b} &:&\qquad \alpha \left| \psi \right\rangle +\beta
\left| \psi ^{\prime }\right\rangle , \\
\left| 1\right\rangle _{b} &:&\qquad \alpha \left| \psi \right\rangle -\beta
\left| \psi ^{\prime }\right\rangle 
\end{eqnarray}
Substituting Eqs. (9) and (10) into Eqs. (12) and (13), and normalizing
them, we have 
\begin{eqnarray}
\left| 0\right\rangle _{b} &:&\qquad \alpha \left( A\left| 0\right\rangle
_{a}+B\left| 1\right\rangle _{a}\right) +\beta \left( A\left| 1\right\rangle
_{a}-B\left| 0\right\rangle _{a}\right) , \\
\left| 1\right\rangle _{b} &:&\qquad \alpha \left( A\left| 0\right\rangle
_{a}+B\left| 1\right\rangle _{a}\right) -\beta \left( A\left| 1\right\rangle
_{a}-B\left| 0\right\rangle _{a}\right) ,
\end{eqnarray}
where the coefficients $A$ and $B$ are 
\begin{eqnarray}
A &=&\frac{\prod_{i=1}^{m}\cos \theta _{i}\prod_{i=m+1}^{n}\sin \theta _{i}}{%
\sqrt{\left( \prod_{i=1}^{m}\cos \theta _{i}\prod_{i=m+1}^{n}\sin \theta
_{i}\right) ^{2}+\left( \prod_{i=1}^{m}\sin \theta _{i}\prod_{i=m+1}^{n}\cos
\theta _{i}\right) ^{2}}} \\
B &=&\frac{\left( -1\right) ^{m}\prod_{i=1}^{m}\sin \theta
_{i}\prod_{i=m+1}^{n}\cos \theta _{i}}{\sqrt{\left( \prod_{i=1}^{m}\cos
\theta _{i}\prod_{i=m+1}^{n}\sin \theta _{i}\right) ^{2}+\left(
\prod_{i=1}^{m}\sin \theta _{i}\prod_{i=m+1}^{n}\cos \theta _{i}\right) ^{2}}%
}.
\end{eqnarray}
Noting that $A$ and $B$ satisfy $A^{2}+B^{2}=1,$ we can define $A=\cos \phi $
and \ $B=\sin \phi .$ Thus, Eq. (14) and Eq. (15) can be written as 
\begin{eqnarray}
\left| 0\right\rangle _{b} &:&\qquad \alpha \left( \cos \phi \left|
0\right\rangle _{a}+\sin \phi \left| 1\right\rangle _{a}\right) +\beta
\left( -\sin \phi \left| 0\right\rangle _{a}+\cos \phi \left| 1\right\rangle
_{a}\right) , \\
\left| 1\right\rangle _{b} &:&\qquad \alpha \left( \cos \phi \left|
0\right\rangle _{a}+\sin \phi \left| 1\right\rangle _{a}\right) -\beta
\left( -\sin \phi \left| 0\right\rangle _{a}+\cos \phi \left| 1\right\rangle
_{a}\right) 
\end{eqnarray}
where 
\begin{equation}
\phi =\tan ^{-1}\frac{B}{A}=\tan ^{-1}\left[ \left( -1\right)
^{m}\prod_{i=1}^{m}\tan \theta _{i}\prod_{i=m+1}^{n}\cot \theta _{i}\right] 
\end{equation}
\ \ \ \ The above results (18-19) imply that when the qubit $b$ is measured
in the $\left| 0\right\rangle $ state$,$ the resulting state (18) of qubit $a
$ is the same as the state $R\left( \phi \right) \left( \alpha \left|
0\right\rangle _{a}+\beta \left| 1\right\rangle _{a}\right) $, i.e., the
state created by Alice directly performing a rotation operation $R\left(
\phi \right) $ on the initial state $\alpha \left| 0\right\rangle +\beta
\left| 1\right\rangle $ of qubit $a$. This means that in the case when the
qubit $b$ is measured in the $\left| 0\right\rangle ,$ the above $n$ users
apply a rotation operation $R\left( \phi \right) $ to a distant qubit $a$
through their local operations. From Eq. (20), the rotation angle $\phi $
depends on each user's rotation angle. On the other hand, when the qubit $b$
is measured in the $\left| 1\right\rangle $ state, the resulting state (19)
of qubit $a$ is the same as the state created by Alice directly performing a
Pauli rotation $\sigma _{z}$ (i.e., a phase-flip operation) and then a
rotation operation $R\left( \pi +\phi \right) $ on the initial state $\alpha
\left| 0\right\rangle +\beta \left| 1\right\rangle $ of qubit $a.$ This
shows that in the case when the qubit $b$ is measured in the $\left|
1\right\rangle ,$ the above $n$ users apply a rotation operation $R\left(
\pi +\phi \right) $ together with a Pauli rotation $\sigma _{z}$ to a
distant qubit $a.$ Alternatively, for this measurement outcome, \ Alice
could apply a $\sigma _{z}$ to the qubit $a$, with the result that its state
will become $R\left( \pi -\phi \right) \left( \alpha \left| 0\right\rangle
+\beta \left| 1\right\rangle \right).$  In terms of Bloch-sphere rotations, 
the first possibility (state $|0\rangle$) corresponds to a rotation by an angle 
$2\phi$ around the $y$ axis, whereas the second possibility (state $|1\rangle$), 
after Alice applies the $\sigma_z$ operation, is a rotation by an angle 
$2\pi-2\phi$, or, alternatively, a rotation by an angle $2\phi$ in the 
{\it opposite\/} direction. 

In previous work [7], Huelga et al. have shown that if Bob wants to perform an
arbitrary unitary remote operation on Alice' particle, Bob and Alice need to
share two EPR pairs (i.e., two e-bits), and also two classical bits are
required. More recently [8] they have also considered the resources needed
to perform restricted sets of operations (in particular, Bloch-sphere rotations),
and found that for two classes of operations, namely, rotations (by any angle) around the 
$z$ axis, or rotations by $\pi$ degrees around any axis lying on the $x-y$ plane, 
only one bit of entanglement and 2 classical bits (one in each direction) are
needed, if one knows ahead of time what kind of transformation the other party
is trying to implement.  

In our case, if we restrict ourselves to just two parties, Alice and Bob, we have
$n=1$, and the GHZ state is just a two-qubit Bell state.  Inspection shows that,
if Alice knows the result of Bob's measurement on his qubit, she can 
(by applying suitable local operations, such as phase and/or bit flips) put her qubit
in a state which can be written compactly as $\exp(\pm i\sigma_y\theta)|\rangle_a$,
where $\theta$ is the rotation angle chosen by Bob, $|\rangle_a$ is her qubit's
initial state, and (depending on the result of her own measurement on qubit $b$) 
she will know whether 
the $+$ or the $-$ sign applies in this expression.  In other words, for this special case,
one e-bit and one classical bit are enough to remotely effect a rotation of the right
magnitude, but its direction (clockwise or counterclockwise) is random, depending
on the outcome of Alice's measurement of her qubit $b$. As will be argued in the following 
section, in spite of its randomness, it is possible that this result might be useful in 
some contexts.

\begin{center}
{\bf III. APPLICATION TO SECRET SHARING}
\end{center}

A practical application for the above ``teleporting'' of a rotation may be a
modification of the quantum secret sharing ideas first presented by Hillery
et al. [3]. In their scheme, they show how to obtain quantum secret sharing
by splitting quantum information among several parties, in such a way that
only one of them is able to recover the qubit exactly provided all the other
parties agree to cooperate. However, we note that, in general,
each party still can get partial quantum information in this scheme.
According to the protocols in Ref. [3], when Alice's qubit is 
in the state $\alpha \left| 0\right\rangle +\beta \left| 1\right\rangle $, the 
$n$ parties who receive the information are left sharing an $n$-qubit entangled 
state of the form $\alpha \left| 00...0\right\rangle \pm \beta \left|
11...1\right\rangle $ or $\alpha \left| 11...1\right\rangle \pm \beta \left|
00...0\right\rangle $ (depending on Alice's Bell-state measurement results).
From this one can get the density operator for each qubit $\rho _{1}=\rho
_{2}...=\rho _{n}=\left| \alpha \right| ^{2}\left| 0\right\rangle
\left\langle 0\right| +\left| \beta \right| ^{2}\left| 1\right\rangle
\left\langle 1\right| $ or $\left| \alpha \right| ^{2}\left| 1\right\rangle
\left\langle 1\right| +\left| \beta \right| ^{2}\left| 0\right\rangle
\left\langle 0\right| .$ This expression shows that, even without cooperating with
others, each party can still get some amplitude information about Alice's
qubit, although neither of them can independently recover the full original state. 

In the following, we will present a new scheme for quantum secret
sharing, which we call ``receiver-encoded'' secret sharing. As we will show,
the present scheme is actually not based on splitting quantum information; rather,
it works essentially through each receiver ``encoding Alice's message
qubit'' by their respective rotation angles and then by Alice sending her
rotated qubit to one of the parties.  A nice property of our scheme is that,
if the method presented in the previous section is used, the receivers can perform
the remote rotation of Alice's qubit without having access to {\it any\/} of the
information contained in it (whether individually or jointly); 
this can be seen immediately from the fact that, when
the overall state of the system is given by (3) above, the reduced density operator 
for the $n$ receivers is just $|00\ldots 0\rangle\langle 00\ldots 0| + 
|11\ldots 1\rangle\langle 11\ldots 1|$, independent of $\alpha$ and $\beta$.

Suppose then that the $n$ receivers follow the procedure in the previous section
to remotely rotate Alice's qubit $a$, after which she sends it to one of them (e.g., Bob). 
The state of the rotated qubit $a$ is given by Eqs. (18) and (19). Noting that the
rotation operator $R\left( \phi \right)$ and the Pauli operator $\sigma _{z}$ are given by 
\begin{equation}
R\left( \phi \right) =\left( 
\begin{array}{cc}
\cos \phi  & \sin \phi  \\ 
-\sin \phi  & \cos \phi 
\end{array}
\right) ,\qquad \sigma _{z}=\left( 
\begin{array}{cc}
1 & 0 \\ 
0 & -1
\end{array}
\right) ,
\end{equation}
we have 
\begin{eqnarray}
R\left( -\phi \right) R\left( \phi \right)  &=&I, \\
\sigma _{z}R\left( -\pi -\phi \right) R\left( \pi +\phi \right) \sigma _{z}
&=&I
\end{eqnarray}
where $I$ is the identity. Thus, if Bob wants to recover the original state $%
\alpha \left| 0\right\rangle _{a}+\beta \left| 1\right\rangle _{a}$ of qubit 
$a$ from Alice$,$ (a) for the case when Alice measures the qubit $b$ in the $%
\left| 0\right\rangle $, he may perform a unitary operation $R\left( -\phi
\right) $ (i.e., the rotation with the angle $-\phi )$ on the qubit $a;$ (b)
for the case when the qubit $b$ is measured in the $\left| 1\right\rangle,$
he can perform a unitary operation $\sigma _{z}R\left( -\pi -\phi \right) $
(i.e., first, a rotation with the angle $-\pi -\phi $, and then a Pauli
rotation $\sigma _{z})$ on qubit $a.$

From the above description, one can see that Bob's recovery operation
depends on Alice's measurement outcomes, and that the angle $\phi$ (given
by Eq. (20)) depends on each receiver's rotation angle and each receiver's
measurement outcome. Thus, in order for Bob to recover the original state of
qubit $a,$ (a) Alice needs to send her measurement outcomes to Bob through a
classical channel, and (b) all the other receivers need to tell Bob their
rotation angles and their measurement outcomes through classical
communications. If any other receiver does not collaborate with Bob, he
has no way to calculate the value of $\phi $ accurately, and thus he can do
no better than to leave the qubit in a random rotated state.

We now need to ask the following question: suppose that we have a qubit and apply
 to it a random rotation; how well, on
the average, have we ``hidden'' the information initially contained in it?
 One way to answer this is to look at the average fidelity between the
rotated qubit and the original one; if the average is $1/2$, then the
rotated qubit might as well be a totally random state; otherwise there is
some ``trace'' of the original state left in the rotated state.

If the initial state is of the general form

\begin{equation}
\left| \psi \right\rangle =\cos \frac{\vartheta }{2}\left| 0\right\rangle
+e^{i\varphi }\sin \frac{\vartheta }{2}\left| 1\right\rangle ,
\end{equation}
the rotated state is of the form 
\begin{equation}
R\left( \phi \right) \left| \psi \right\rangle =\left( \cos \frac{\vartheta 
}{2}\cos \phi -e^{i\varphi }\sin \frac{\vartheta }{2}\sin \phi \right)
\left| 0\right\rangle +\left( \cos \frac{\vartheta }{2}\sin \phi
+e^{i\varphi }\sin \frac{\vartheta }{2}\cos \phi \right) \left|
1\right\rangle
\end{equation}
and the fidelity is 
\begin{equation}
{\cal F}\left( \vartheta ,\varphi ,\phi \right) =\left| \left\langle \psi
\right| R\left( \phi \right) \left| \psi \right\rangle \right| ^{2}=\cos
^{2}\phi +\sin ^{2}\phi \sin ^{2}\vartheta \sin ^{2}\varphi .
\end{equation}
If this is averaged over $\vartheta $, $\varphi $, and $\phi $, assuming
uniform (random) distributions, the result is $5/8=0.625$, greater than the
relative fidelity between the initial state and the totally random state. On
the other hand, for the special class of qubits for which $\varphi =0$, we
have

\begin{equation}
{\cal F}\left( \vartheta ,\varphi ,\phi \right) =\cos ^{2}\phi
\end{equation}
and this clearly averages to $1/2.$

Note that for the qubit rotated by the many receivers, as above, the
distribution of the overall rotation angle $\phi $ is not uniform, but one
should still have $\left\langle \cos ^{2}\phi \right\rangle =\left\langle
\sin ^{2}\phi \right\rangle =1/2,$ since, for instance, $\cos ^{2}\phi
=1/2+\left( \cos 2\phi \right) /2$ and the average of $\cos 2\phi $ is zero
for any distribution of $\phi $ (between $0$ and $\pi $) which is symmetric
around $\pi /2$, which will be the case for the distribution of the overall
rotation angle $\phi $ (given by Eq. (20)) if all the distributions for $%
\theta _{1}$, $\theta _{2},...$ have the same symmetry. Thus, for the
special case of qubits of the form (24) with $\varphi =0$, our method
produces a qubit which, on average, bears no more resemblance to the initial
qubit than a totally random state.

For qubits with $\varphi \neq 0,$ as stated above, the situation is
different. In particular, for the special cases $\left| \psi \right\rangle =%
\frac{1}{\sqrt{2}}\left( \left| 0\right\rangle \pm i\left| 1\right\rangle
\right) ,$ the rotation leaves the qubit invariant except for a phase change
(cf. eq. (26) with $\vartheta =\frac{\pi }{2},$ $\varphi =\pm \frac{\pi }{2}$;
this is natural, since these are the eigenstates of $\sigma_y$). 
Hence, these special qubit states are not ``hidden'' at all. We have,
therefore, a ``restricted'' secret sharing scheme, in a sense
complementary to the one of Hillery et al. [3]: in their scheme, the sharers
could get information on the original qubit's amplitude, in ours they could
get information on the phase. The main difference is that in our scheme only
one user actually has a physical qubit (which may in some way be regarded as
safer).

Although we shall not attempt here a comprehensive study of the security of the 
scheme against all possible forms of eavesdropping and/or cheating, we believe
that it is probably quite secure, for several reasons.  First, as
pointed out above, before Alice actually sends her qubit to one of the receivers
they have no information at all, either individually or jointly, 
about its state (assuming that the overall state
of the system is given by Eq. (3)); from here it also follows that an eavesdropper
cannot hope to gain information about Alice's qubit by entangling a particle with
any of the receivers'.  Second, the qubit Alice sends to (say) Bob is basically 
useless (modulo the reservations just discussed above) without the classical information
possessed individually by all the ``receivers;'' hence, even if Eve were to intercept the 
qubit intended for Bob, and replace it by a fake, and somehow eavesdropped on
the (classical) communication channels through which all the other parties disclose
to Bob their rotation angles and measurement outcomes, she would still not be able
to recover the qubit's original state without access to Bob's own information (his
rotation angle and measurement outcome), which he does not have to send to anybody, and
hence may be considered secure.  It is conceivable that an eavesdropper might 
get partial information on the rotation angles and measurement outcomes of all the
receivers by entangling enough particles with their respective qubits, but 
presumably such entanglement could be detected by
tests conducted on ``sample'' GHZ states initially shared by Alice and the other
parties, as discussed by Hillery et al. [3].

\begin{center}
{\bf IV. CONCLUSION}
\end{center}

We have presented a (probabilistic) method to teleport a certain class of rotations and
proposed a possible application to a ``restricted'' quantum secret-sharing
scheme. A special feature of our secret-sharing concept is that only one
recipient actually gets a qubit of quantum information; all the other
parties have only the classical information of their rotation angles, known
only to themselves. Thus the original quantum information is not really
split into ``shares'': the quantum channel, consisting of the shared GHZ
state, is used only for the receivers to rotate the original qubit by their
local operations known only to themselves (and without gaining any information on the
original qubit state in the process). Although our scheme is less general
than, for instance, the threshold schemes of Ref. [5], we believe it is of
some interest nonetheless, especially because of its relatively
straightforward nature.

This work has been supported in part by the National Security Agency (NSA)
and Advanced Research and Development Activity (ARDA) under Army Research
Office (ARO) contract number DAAD19-99-1-0118; and by the National Science
Foundation under grant PHY-9802413.

\end{document}